\documentstyle[12pt]{article}

\input epsf.tex
\def\DESepsf(#1 width #2){\epsfxsize=#2 \epsfbox{#1}}

minus.5em \abovedisplayshortskip=2mm plus.2em minus.4em
\def\title#1#2#3#4#5{
\begin{center} \begin{tabular}[t]{l} #1 \end{tabular} \hfill
\begin{tabular}[t]{r} #2 \end{tabular} \\[1cm] {\LARGE\bf #3} \\[.5in] {#4{}}
\end{center} \vfill \centerline{{\large ABSTRACT}} {\nopagebreak
\noindent\begin{quotation}\noindent {\small #5}

\end{quotation}} \vfill \newpage
\def\thefootnote{\sharp\arabic{footnote}}}

\begin {document}

\thispagestyle{empty}
\title {July 1995} {\bf OSU-298\\ \bf UCRHEP-T139} {Effects of supersymmetric
grand unification scale physics on
$\Gamma \left( b\rightarrow s\gamma\right)$} { {\bf B. Dutta$^{\dagger} $ and
E.
Keith $^{\ast} $ }\\
$^{\dagger} $ Department of Physics\\Oklahoma State University\\ Stillwater, OK
74078\\$^{\ast} $ Department of Physics \\University of California\\Riverside,
CA 92521} {Although calculations of the $b\rightarrow s\gamma$ rate in
supersymmetric grand unified models have always either ignored the gluino
mediated contribution or found it to be negligible, we show that taking
universal supersymmetry breaking masses at the Planck scale, rather than at the
gauge unification scale as is customary, leads to the gluino contribution being
more significant and in fact sometimes even larger than the chargino mediated
contributions when $\mu >0$ and $\tan{\beta}$ is of order 1. The impact is
greatest felt when the gluinos are relatively light. Taking the universal
boundary condition at the Planck scale also has an effect on the chargino
contribution by increasing the effect of the wino and higgsino-wino mediated
decays. The neutralino mediated contribution is found to be enhanced, but
nevertheless it remains relatively insignificant.}

The flavor changing decay $b\rightarrow s\gamma$ is often an important test of
new physics because it is rapid enough to be experimentally observable although
it appears first at the one loop level in the standard model (SM), thus
allowing
new physics to add sizeable corrections to it. For example, the decay is useful
to limit parameter space in the minimal supersymmetric standard model (MSSM)
\cite{[AM],[NO],[RB],[VB],[RA],[JH],[MD],[GN],[FB],[NL],[SV]}. This is an
especially useful tool if certain constraints have already been placed on the
MSSM. Since the decay vanishes in the limit of unbroken supersymmetry, the
relevant constraints pertain to the terms in the Lagrangian that softly break
supersymmetry (SUSY). The general soft SUSY breaking scalar interactions for
squarks and sleptons in the MSSM are of the following form: \begin{eqnarray}
V_{\rm soft}&=&Q({\bf A}_U{\bf \lambda}_U)U^cH_2+Q({\bf A}_D{\bf
\lambda}_D)D^cH_1+L({\bf A}_E{\bf
\lambda}_E)E^cH_1 +
\rm{h.c.}\nonumber\\&+& Q^\dagger {\bf m}_Q^2Q+U^{c\dagger}{\bf
m}_U^2U^c+D^{c\dagger}{\bf m}_D^2D^c + L^\dagger {\bf m}_L^2L+E^{c\dagger}{\bf
m}_E^2E^c\, , \end{eqnarray} where $Q$ and $L$ are squark and slepton doublets
and $U^c$, $D^c$, and $E^c$ are squark and slepton singlets. It is most
frequently assumed that soft breaking operators are induced by supergravity,
and
that these operators have a universal form which is generation symmetric and CP
conserving. Usually this assumption is coupled with that of grand unification
to
provide further constraints on the model's parameters. For purposes of
simplicity the universal boundary condition is traditionally taken at the grand
unification scale, even though the soft-breaking operators would be present up
to near the Planck scale. The universal soft SUSY-breaking boundary condition
is
described by the following parameters: the scalar mass $m_0$, the gaugino mass
$m_{1/2}$, and the trilinear and bilinear scalar coupling parameters $A$ and
$B$, respectively.

Despite convention, some recent papers \cite{[NP],[LJ],[SD],[AS]} have
demonstrated important implications of taking the boundary conditions at the
Planck scale $M_P$ and evolving them down to the scale $M_G$ where grand
unification is broken. The important difference between taking the universal
boundary condition at the Planck scale versus at the GUT scale is due to that
the top quark mass is known to be large ($\sim 174\rm{GeV}$) and grand
unification causes some fields to feel the effects of the top coupling by
unifying them into the same multiplet with the top. For example in SU(5) grand
unification, $Q$, $U^c$, and
$E^c$ together transform as the $\bf{10}$-representation. Taking the universal
boundary condition at the GUT scale is ignoring the fact that the soft-breaking
parameters run above the GUT scale. One may at first think that any effects of
grand unification would be supressed by powers of $1/M_G$, but it has been
demonstrated that such effects rather depend on $\ln{(M_P/M_G)}$ \cite{[KO]}.
One surprising result, which is given in refs. \cite{[LJ],[AS]}, is that the
predicted rate for the lepton flavor violating decay $\mu\rightarrow e\gamma$
may be only one order of magnitude beneath current experimental limits. This is
found even in SU(5) grand unification, where only neutralinos are available to
mediate the decay at the one-loop order. Although due to the soft-breaking
masses being assumed flavor blind and the unitarity of the CKM mixing matrix
one
might naively expect the partial width to vanish, this does not happen because
operators with strength of the top coupling cause the third generation scalar
fields contained in the ${\bf 10}$-representation, in SU(5), to be considerably
lighter than the corresponding fields of the other two generations \cite{[LJ]}.
Below the grand unification scale, the only squarks effected by the top Yukawa
coupling are the top singlet and the third generation SU(2)$_L$ squark doublet.

We will apply these facts to demonstrate that in some of the same regions of
parameter space where chargino corrections are important, one may also expect
sizeable corrections from gluinos. Previous calculations of $b\rightarrow
s\gamma$ in the MSSM have routinely either ignored the contributions mediated
by
gluinos
\cite{[RB],[VB],[RA],[JH],[MD],[GN],[NL]} or found them to be less important
\cite{[AM],[NO],[FB],[SV]} than we will find them to be when we take the
universal SUSY soft breaking boundary condition at the Planck scale rather than
at the GUT breaking scale, as is cutomary. This is due to the fact, as
explained above, that taking the universal boundary condition at the Planck
scale
leads to a different low energy SUSY spectrum than taking it at the GUT
breaking
scale does.
 For some parameter space, not only is the
contribution from the gluinos important, but it is also comparable to
that of the charginos. The fact that the top squark soft breaking masses are
lighter when the universal boundary condition is taken at the Planck scale
rather than the grand unification scale is, of course, likewise felt by the
wino
and higgsino-wino mediated decays.

The standard model amplitude for $b\rightarrow s\gamma$ has been derived in
refs.
\cite{[TI],[NG]}, and expressions for the additional MSSM amplitudes have been
derived in refs. \cite{[AM],[NO],[RB]}, with ref. \cite{[AM]} containing the
first and most complete derivation. The leading-order QCD corrected
 partial width is given by its ratio to the inclusive
semi-leptonic decay width in the following form \cite{[AJ]}: \begin{eqnarray}
{\Gamma \left( b\rightarrow s\gamma\right)\over \Gamma \left( b\rightarrow
ce\nu
\right) }={6\alpha \over \pi g(z)}{\left| V_{ts}^*V_{tb}\right|^2\over
\left| V_{cb}\right|^2}\left| c_7\left( m_b\right)\right|^2\, , \end{eqnarray}
where $\alpha$ is the electromagnetic coupling, $g(z)
=1-8z^2+8z^6-z^8-24z^4\ln{z}$ with $z=m_c/m_b=0.316\pm 0.013$ is the
phase-space
factor, and the inclusive semi-leptonic branching ratio is $BR ( b\rightarrow
ce\nu )=0.107$. The ratio of the CKM matrix entries, for which we will use the
experimental mid-value, is
$\left| V_{ts}^*V_{tb}\right|^2/
\left| V_{cb}\right|^2=0.95\pm 0.04$. The QCD corrected amplitude $c_7(m_b)$ is
given as
\begin{eqnarray} c_7\left( m_b\right) =\eta^{16/23} \left[ c_7\left( M_W\right)
-{8\over 3}c_8\left( M_W\right) \left[ 1-\eta^{-2/23}\right]\right]
+\sum_{i=1}^8{a_i\eta^{b_i}} \, , \end{eqnarray} with $a_i$ and $b_i$ being
given in ref. \cite{[AJ]}, $\eta =\alpha_s(M_W) /\alpha_s(m_b)$, for which we
will use $\eta =0.548$. The terms
$c_7(M_W)$ and
$c_8(M_W)$ are respectively $A_\gamma$, the amplitude for $b\rightarrow
s\gamma$
evaluated at the scale $M_W$ and divided by the factor $A_\gamma^0\equiv
2G_F\sqrt{\alpha / 8\pi^3}V_{ts}^*V_{tb}m_b$ and $A_g$, the amplitude for
$b\rightarrow sg$ divided by the factor $A_\gamma^0 \sqrt{\alpha_s /\alpha}$.
The effective interactions for $b\rightarrow s\gamma$ and $b\rightarrow sg$ are
given by
\begin{eqnarray} L_{\rm eff}={A_\gamma^0 \over 2}\left( A_\gamma
\overline{s}\sigma^{\mu\nu} P_RbF_{\mu\nu}+A_g\overline{s}\sigma^{\mu\nu}
P_RbG_{\mu\nu}\right) +{\rm h.c.}\, .
\end{eqnarray} In calculating $c_7(M_W)$ and
$c_8(M_W)$, we will however use the conventional approximation of taking the
complete MSSM to be the correct effective field theory all the way from the
scale $M_G$ down to
$M_W$, and hence ignore threshold corrections.

As previously stated, normally
$A_\gamma$ is taken to be approximately the sum of $A_\gamma^W$,
$A_\gamma^{H^-}$, and $A_\gamma^{\tilde{\chi}^-}$. In such a case, the charged
Higgs contribution adds constructively to the SM amplitude. On the other hand,
the chargino amplitude may combine either constructively or destructively with
the other two, and in some cases may even cancel the charged Higgs amplitude.
Even though the squarks strongly couple to the gluino, the contribution from
the
gluino mediated diagrams are considered negligible because the three
generations
of down squarks $\tilde{d}_{iL}$ belonging to $\tilde{Q}_{Li}$ are
conventionally assumed to have degenerate soft-breaking masses at the GUT
breaking scale. However, the mass parameter $m^2_{\tilde{Q}_{3L}}$ is reduced
by
a small amount relative to $m^2_{\tilde{Q}_{iL}}$ for the first two generations
in running the mass parameters down from the GUT scale, and b-squark mass
matrix
has off-diagonal entries proportional to $m_b$ as given in the following
equation: \begin{eqnarray} m^2_{\tilde{b}}=\left( \matrix{
m^2_{\tilde{Q}_{L3}}+m_b^2-{1\over 6}\left( 2M_W^2+M_Z^2\right) \cos{2\beta} &
m_b\left( A_b+\mu \tan{\beta}\right) \cr m_b\left( A_b+\mu \tan{\beta} \right)
&
m^2_{\tilde{b}_{R}}+ m_b^2+{1\over 3}\left( M_W^2-M_Z^2\right) \cos{2\beta}\cr
}\right)\, ,
\end{eqnarray} where $\tan{\beta}=v_2/v_1$ is the ratio of Higgs vacuum
expectation values and $\mu$ is the coefficient of the Higgs superpotential
interaction $\mu H_1H_2$. These two effects make the b-squark eigenvalues
somewhat different from the down squark masses of the other two generations,
however the total effect is insignificant compared to the mass splitting that
takes place in the stop sector due to the size of the top quark mass. (See, for
example, Fig. 8 in ref. \cite{[SUSY]}.) For this reason, and because the
chargino contribution includes an often highly significant higgsino mediated
decay, the chargino contribution to
$b\rightarrow s
\gamma$ is found to be very important for some regions of parameter space,
while
the gluino and neutralino contributions are conventionally either found or
assumed to be of little signifigance when the universal boundary condition is
taken at the scale
$M_G$.

Now, we will perform the calculation with the universal boundary condition
taken
at the Planck scale and run the soft breaking parameters from there down to the
weak scale. In the following discussion we will use the one loop
renormalization
group equations for the gauge couplings, top Yukawa coupling, and soft breaking
masses (See refs. \cite{[NP],[KIN]}). We will also use the exact analytic
solutions, in the form derived in ref. \cite{[AS]}, to these one loop
equations.
We will use the conventions for the sparticle mass matrices and the trilinear
coupling parameter $A_i$ as found in ref. [SUSY]. ($A_i\rightarrow -A_i$ in the
RGEs and RGE solutions of ref. \cite{[AS]}.) We will use $\alpha_s (M_Z)=0.12$.
For the purpose of illustration, we will consider the specific case where the
Planck scale trilinear scalar coupling $A_0=0$, $\tan{\beta}=1.5$,
$\lambda_t(M_G) =1.4$, and the grand unification model is the minimal SUSY
SU(5)
model. In another paper \cite{[DDK]}, we will look at larger regions of
parameter
space, including large $\tan{\beta}$, for SO(10) grand
unification. If
$\lambda_t(M_G)$ is reduced significantly then so also would be the effects we
are discussing. For our chosen values of $\alpha_s (M_Z)$, the $M_G$ scale top
coupling, and $\tan{\beta}$, the top quark pole mass is 168 GeV and the SM
branching ratio is about $3.0\cdot 10^{-4}$. For larger $\tan{\beta}$ the
gluino
and neutralino contribution would be greatly increased \cite{[NO]}, but at the
same time this would tend to occur for sparticle masses where the chargino
contribution to
$b\rightarrow s\gamma$ is large enough to rule out the region of parameter
space.

Since the off diagonal terms in the b-squark mass matrix are much smaller than
the diagonal ones and give relatively only a small contribution to the mass
splitting between the b-squark mass eigenstates, we choose for simplicity to
take the b-squark mass eigenstates to be the soft breaking masses. (See ref.
\cite{[SD]}.) The two types of diagrams that contribute to $b\rightarrow
s\gamma$ and
$b\rightarrow sg$ in SU(5) with $\tilde{d}_{iL}$ running in the internal loop
are shown in Fig. 1. The internal fermion line represents either a gaugino or a
neutralino propagator. To derive the contributions to $A_\gamma$ or $A_g$, one
must sum the graphs with an external photon or gluon, respectively, attached in
all possible ways. It is possible and simplest to work in a basis in which
${\bf\lambda_U}$, the soft breaking squark masses and the trilinear couplings
$A$ are always diagonal in generation space \cite{[SD]}. The masses of the
first
two generations of squarks $\tilde{d}_{iL}$ are essentially equal to their soft
breaking masses, which receive renormalization effects only from gaugino loops,
and are hence degenerate. The soft breaking mass of $\tilde{b}_L$ is much
smaller than that of the other two generations, as we will see, since the
$\tilde{b}_L$ belongs to the same multiplet as the top above $M_G$. Noting that
there is no mixing at the
$b_R$-$\tilde{b}_R$-gaugino vertex introduced by SU(5) grand unification and
using the unitarity of the CKM matrix ${\bf V}$, one may express the
contributions to
$A_\gamma$ by the gluinos
\cite{[AM],[NO],[AS]} as follows: \begin{eqnarray}A_\gamma^{\tilde{g}}&=&
-C(R)e_DM_W^2{\alpha_s\over \alpha_2} \{ {g_1\left(
m_{\tilde{Q}_{3L}}^2/M_3^2\right) - g_1\left(
m_{\tilde{Q}_{1L}}^2/M_3^2\right)\over M_3^2}
\nonumber\\ &+& \eta_b^{-1}
\left( A_d+\mu \tan{\beta}\right) \left[ G\left(
m_{\tilde{b}_{R}}^2,m_{\tilde{Q}_{3L}}^2\right) - G\left(
m_{\tilde{b}_{R}}^2,m_{\tilde{Q}_{1L}}^2\right)\right] \nonumber\\ &+&
\eta_b^{-1}\left( A_b-A_d\right) G\left(
m_{\tilde{b}_{R}}^2,m_{\tilde{Q}_{3L}}^2\right)\}\, , \end{eqnarray} where
$G$ is given by
\begin{eqnarray} G\left( m_1^2,m_2^2\right) ={1\over M_3}{g_2\left( {m_1^2\over
M_3^2}\right) -g_2\left( {m_2^2\over M_3^2}\right) \over m_1^2-m_2^2}\, ,
\end{eqnarray} with
\begin{eqnarray} g_1(r)&=&{1\over 6(r-1)^4}[2+3r-6r^2+r^3+6r\ln{r}]\, ,\\
g_2(r)&=& {-1\over 2(r-1)^3}[r^2-1-2r\ln{r}]\, , \end{eqnarray}where we have
used $e_D=-1/3$, $C(R)=4/3$, and $\eta_b\equiv m_b(m_b)/m_b(M_W)$, which we
take
to be $\eta_b=1.5$. The analogous expression for neutralinos may be obtained
from the above expressions by working in the neutralino mass eigenbasis and
noting that the first term in Eq. (6) comes from the diagram of Fig. 1a with
the
bino propagator and the other terms come from Fig. 1b with the bino-wino
propagator. (See, for example, ref.
\cite{[AS]}.) We have not included the diagram with the higgsino-wino
propagator. Because neutralinos do not interact with gluons, one finds that
simply
$A_g^{\tilde{\chi}^0}=A_\gamma^{\tilde{\chi}^0}/e_D$. On the other hand because
a gluon can attach to the gluino propagator, one finds
\begin{eqnarray}A_g^{\tilde{g}}=-{C(G)\over C(R)}{1\over
2e_D}A_\gamma^{\tilde{g}} [g_1\rightarrow h_1,g_2\rightarrow h_2]+ {1\over
2e_D}(2-C(G)/C(R))A_\gamma^{\tilde{g}}\, ,\end{eqnarray} with $C(G)=3$ and
\begin{eqnarray}h_1(r)&=&(1-6r+3r^2+2r^3-6r^2\ln{r})/6(r-1)^4\,
,\\h_2(r)&=&-(-1+4r-3r^2+2r^2\ln{r})/2(r-1)^3\, .\end{eqnarray} Notice that the
gluino contribution to $b\rightarrow sg$ can be highly significant.

The mass of $\tilde{Q}_{3L}$ may be expressed in terms of the first generation
squark mass $m^2_{\tilde{Q}_{1L}}$ as
\begin{eqnarray} m_{\tilde{Q}_{3}L}^2=m_{\tilde{Q}_{1L}}^2 -\left(
I_G+{I_Z\over
2}\right)\, , \end{eqnarray} where $I_G=(3/8\pi^2)\int_{M_G}^{M_P}\lambda_t^2
\left( m_H^2+2m^2_{{\bf 10}_3}+A_t^2\right) d\ln{\mu}$, and $I_Z /2$ is the
analogous contribution obtained from running the scale down from $M_G$ to
$M_W$.
The integrals $I_G$ and $I_Z$ may be obtained from the analytic one loop
solutions in terms of $m_{\tilde{Q}_{1L}}^2$ and $M_3(M_W)$ as follows:
\begin{eqnarray} I_G\approx 0.80 m_{\tilde{Q}_{1L}}^2-0.71 M_3^2\, ,\,
I_Z\approx 0.19 m_{\tilde{Q}_{1L}}^2+0.17 M_3^2\, , \end{eqnarray} where we
have
taken $M_P=2.4\cdot 10^{18}\rm{GeV}$. One can also find that the relevant weak
scale trilinear scale couplings are $A_b\approx -1.38 M_3$ and $A_d\approx -
1.55 M_3$. We calculate the parameter
$\mu$ at the tree level and find $\mu^2\approx 1.0 m^2_{\tilde{Q}_{1L}} -0.038
M_3^2-4200 {\rm GeV}^2$.

To illustrate the relative sizes of the seperate contributions to the
$b\rightarrow s\gamma$ amplitude, we plot \begin{eqnarray} r_{A_i}\equiv
A^{i}_{\gamma}/A^W_{\gamma}
\end{eqnarray} versus gluino mass for different values of $m_{\tilde{Q}_{1L}}$
in Fig. 2 for the case $\mu >0$. Notice that the gluino
contribution to
$A_\gamma$ is sometimes comparable to that of the charginos. This happens when
the gluino mass is light.
 Note
that the neutralino contribution is insignificant as usual. In Fig. 3, we plot
the resulting branching ratio for
$b\rightarrow s\gamma$. The dotted line corresponds to the calculation
neglecting gluino and neutralino mediated contributions, while the solid line
represents the full calulation. In both cases, we have included the SM, charged
Higgs and chargino corrections as found in \cite{[RB]}, which work very well
for
low $\tan{\beta}$. When the glunino mass is 150 GeV, the gluino contribution
can
increase the branching ratio by as much as about 20-percent. This happens when
there is strong destructive interference between the charged Higgs and chargino
mediated amplitudes, and the gluino mediated amplitude then contributes to the
branching ratio mainly through its cross term with the SM amplitude. In Fig. 4
and 5, we show the analogous situation with the universal boundary condition
taken at the GUT scale. Notice that when gluino masses are light, the gluino
contributions are about one-third the size as when the boundary condition is
taken at the Planck scale. The effect of taking the universal boundary
condition
at the Planck scale has only a small effect on the total chargino contribution
for the parameter space shown here, however we find that for other nearby
regions, for example with
$\tan{\beta}=2$, the effect of making the chargino contribution more positive,
but smaller in magnitude, is more appearant. In Fig. 6a and 6b, we plot the
branching ratio as a function of the $M_G$ scale gaugino mass
$M_{5G}$ for curves of constant $m_0$ for the cases where the universal
boundary
condition is taken at the Planck scale and at the scale $M_G$, respectively.
Notice that for the curves with $m_0>0$, the branching ratios for the complete
calculation in the two cases differ by about 10-percent when $M_{5G}=$60 GeV,
which corresponds to a not very light gluino mass of about 170 GeV. When $\mu
<0$ and
$\tan{\beta}$ is of order 1, we find the contribution to be much less important
due to a strong destructive interference between the two diagrams in Fig. 1.

In conclusion, if one is to calculate the decay rate for the flavor changing
process $b\rightarrow s\gamma$ in a SUSY GUT with SUSY breaking communicated by
gravity above the GUT breaking scale in the form of soft breaking mass terms,
it
is essential to include the GUT scale renormalization group effects. An
important result of including these renormalization effects is that the gluino
contribution to the decay rate can now no longer be neglected when the glunino
mass is relatively light.

We thank N.G. Deshpande, E. Ma, S. Nandi, and A. Soni. This was work was
supported by Department of Energy grants DE-FG03-94ER 40837 and DE-FG02-94ER
40852.

\vfill
\newpage
\leftline{{\Large\bf Figure captions}}

\begin{itemize}
\item[Fig. 1~:] {{The two types of diagrams with $\tilde{d}_{iL}$ running in
the
internal loop that can contribute signifigantly to $b_R\rightarrow s_L\gamma$
and $b_R\rightarrow s_Lg$. One must sum the graphs with an external photon or
gluon attached in all possible ways.}\label{fig1}}

\item[Fig. 2~:] {{Plots of $r_{A_i}\equiv A^{i}_{\gamma}/A^W_{\gamma}$ versus
gluino mass for different values of
$m_{\tilde{Q}_{1L}}$ for the case
$\mu >0$ and
$\tan{\beta} =1.5$ with the universal boundary condition taken at the scale
$M_P$. The curves correspond to squark masses
$m_{\tilde{d}_{L}}=$200, 300, 400, and 500 GeV. The gluino masses for each
curve
range from 150 GeV to the corresponding value of $m_{\tilde{d}_{L}}$. For
example, $m_{\tilde{d}_{L}} =$200 GeV corresponds to the curve for which the
gluino mass ranges from 120 Gev to 200 GeV. Figs. 2a, 2b , 2c, and 2d
correspond
to the charged Higgs, chargino, gluino, and neutralino contributions,
respectively.}\label{fig2}}

\item[Fig. 3~:] {{Plots of the branching ratio of $b\rightarrow s\gamma$ for
the
case of Fig. 2. The solid lines represent the calculation including SM, charged
Higgs, chargino, gluino, and neutralino contributions. The dashed lines
represent the calculation using only the SM, charged Higgs, and chargino
contributions. The curves represent the same squark masses as have been used in
Fig. 2.}\label{fig3}}

\item[Fig. 4~:] {{Same as Fig. 2, but with the universal boudary condition
taken
at the $M_G$ scale.}\label{fig4}}

\item[Fig. 5~:] {{Same as Fig. 3, but with the universal boudary condition
taken
at the $M_G$ scale as in Fig. 4.}\label{fig5}}

\item[Fig. 6~:] {{Plots of the branching ratio for the case of $\mu >0$ and
$\tan{\beta} =1.5$ as a function of the
$M_G$ scale gaugino mass $M_{5G}$ for curves of constant $m_0$. Fig. 6a
corresponds to the universal boundary condition taken at the Planck scale. Fig.
6b corresponds to the universal boundary condition taken at the GUT breaking
scale. The solid lines represent the calculation including SM, charged Higgs,
chargino, gluino, and neutralino contributions. The dashed lines represent the
calculation using only the SM, charged Higgs, and chargino contributions. The
curves represent, in descending order in the two plots, $m_0 =$0, 250 GeV, and
500 GeV.}\label{fig6}}

\end{itemize}

\vfill
\begin{figure}[htb]
\centerline{ \DESepsf(dess.epsf width 15 cm) }
\smallskip
\nonumber
\end{figure}

\begin{figure}[htb]
\centerline{ \DESepsf(dess1.epsf width 15 cm) }
\smallskip
\nonumber
\end{figure}

\begin{figure}[htb]
\centerline{ \DESepsf(dess2.epsf width 15 cm) }
\smallskip
\nonumber
\end{figure}

\newpage
\begin{thebibliography}{[001]}



\bibitem{[AM]} S. Bertolini, F. Borzumati, A. Masiero, and G. Ridolfi, Nucl.
Phys.
\underline{B353}, 591 (1991).

\bibitem{[NO]} N. Oshimo, Nucl. Phys.
\underline {B404}, 20 (1993).

\bibitem{[RB]} R. Barbieri and G. F. Giudice, Phys. Lett. \underline{B309},
86(1993)

\bibitem{[VB]} V. Barger, M. Berger, P. Ohmann, and R. J. N. Phillips, Phys.
Rev. \underline{D51}, 2438 (1995), and references therein.

\bibitem{[RA]} R. Arnowitt and Pran Nath, CTP-TAMU-65/94, NUB-TH-3111/94.

\bibitem{[JH]} J. L. Hewett, SLAC-PUB-6521.

\bibitem{[MD]} M. A. Diaz, Phys. Lett. \underline{B322}, 207 (1994).

\bibitem{[GN]} R. Garisto and J. N. Ng, Phys. Lett. \underline{B315}, 372
(1993).

\bibitem{[FB]} F. Borzumati, Z. Phys.
\underline{C63}, 291 (1994).

\bibitem{[NL]}J. L. Lopez, D. V. Nanopoulos, and K. Yuan, Phys. Lett.
\underline{B267}, 219 (1991); S. Kelly, J. L. Lopez, D. V. Nanopoulos, H. Pois,
and K. Yuan, Phys. Rev. \underline{D47}, 2461 (1993).

\bibitem{[SV]}S. Bertolini and F. Vissani, SISSA 40/94/EP.


\bibitem{[NP]} N. Polonsky and A. Pomarol, Phys. Rev. Lett. \underline{73},
2292
(1994); N. Polonsky and A. Pomarol, UPR-0627T.

\bibitem{[LJ]} R. Barbieri and L. J. Hall, Phys. Lett. \underline{B338} 212,
(1994).

\bibitem{[SD]}S. Dimopolous and L. J. Hall,Phys. Lett. \underline{B344}, 185
(1995).

\bibitem{[AS]} R. Barbieri, L. J. Hall, and A. Strumia, Nucl. Phys.
\underline{B445} 219 (1995).

\bibitem{[KO]} L. J. Hall, V. A. Kostelecky, and S. Raby, Nucl.
Phys.\underline{B267} 415, (1986).

\bibitem{[TI]} T. Inami and C. S. Lim, Prog. Theor. Phys. \underline{65}, 297
(1981).

\bibitem{[NG]} N. G. Deshpande and G. Eilam, Phys. Rev. \underline{D26}, 2463
(1982).


\bibitem{[AJ]} A. J. Buras, M. Misiak, M. Munz, and S. Pokorski, Nucl. Phys.
\underline{B424}, 374 (1994).


\bibitem{[SUSY]}V. Barger, M. Berger, and P. Ohmann, Phys. Rev. \underline
{D49}
4908 (1994).

\bibitem{[KIN]}K. Inoue, A. Kakuto, H. Komatsu, and S. Takeshita, Prog. Theo.
Phys. \underline {68} 927 (1982).

\bibitem{[DDK]} B. Dutta, E. Keith, and T. V. Duong, Univ. of Ca., Riverside
preprint UCRHEP-T154 and University of Oregon preprint OITS-591 (1995).
\end{thebibliography}
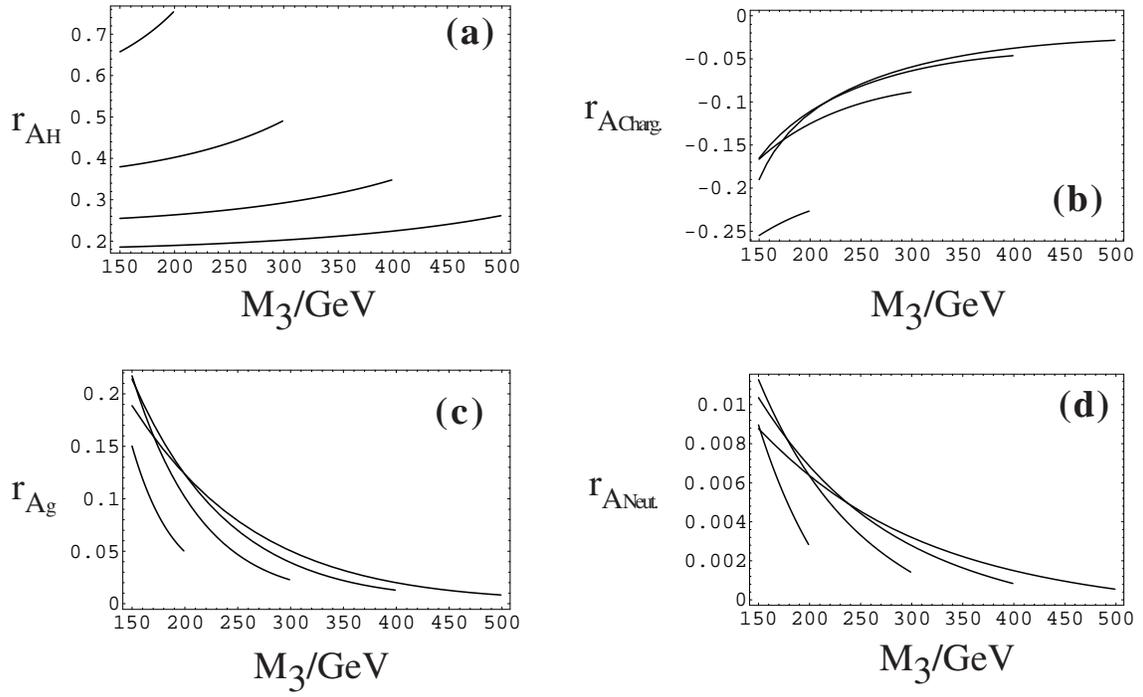
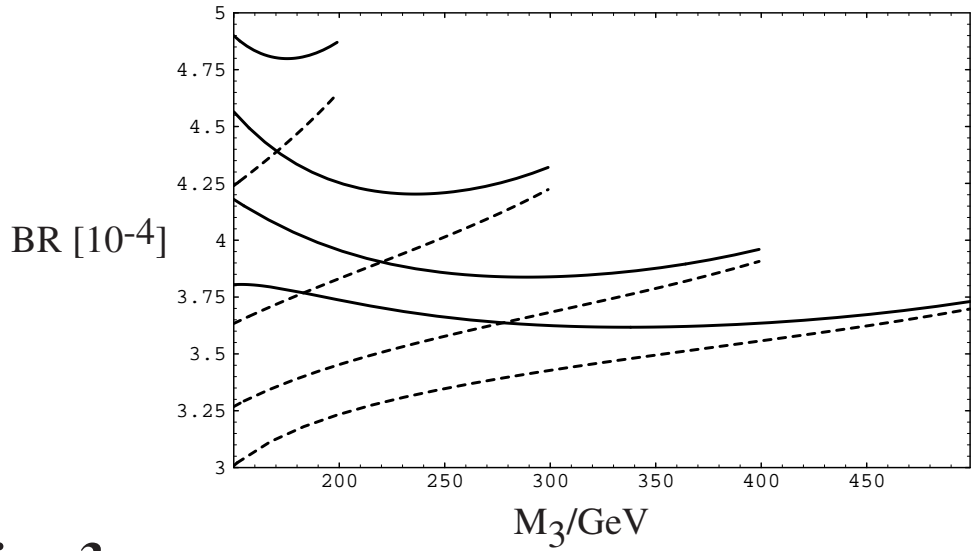
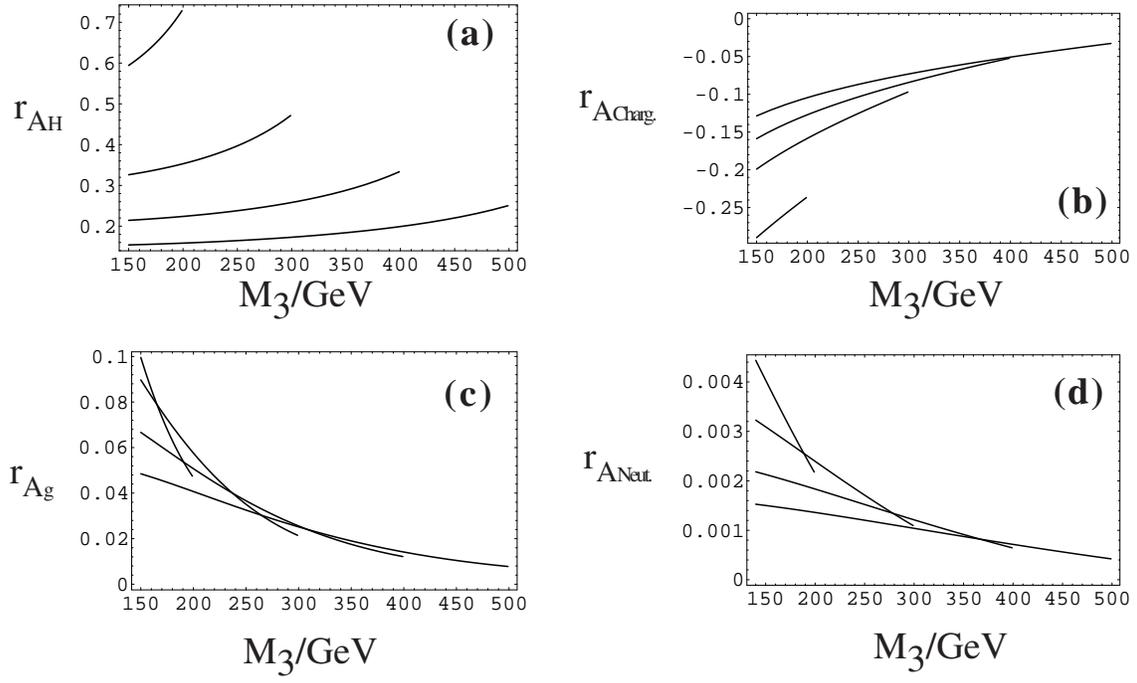
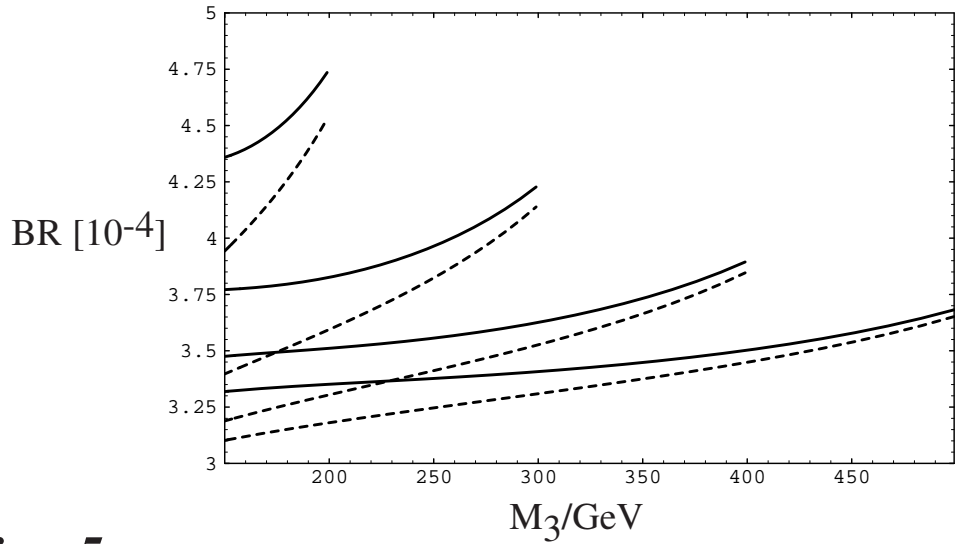
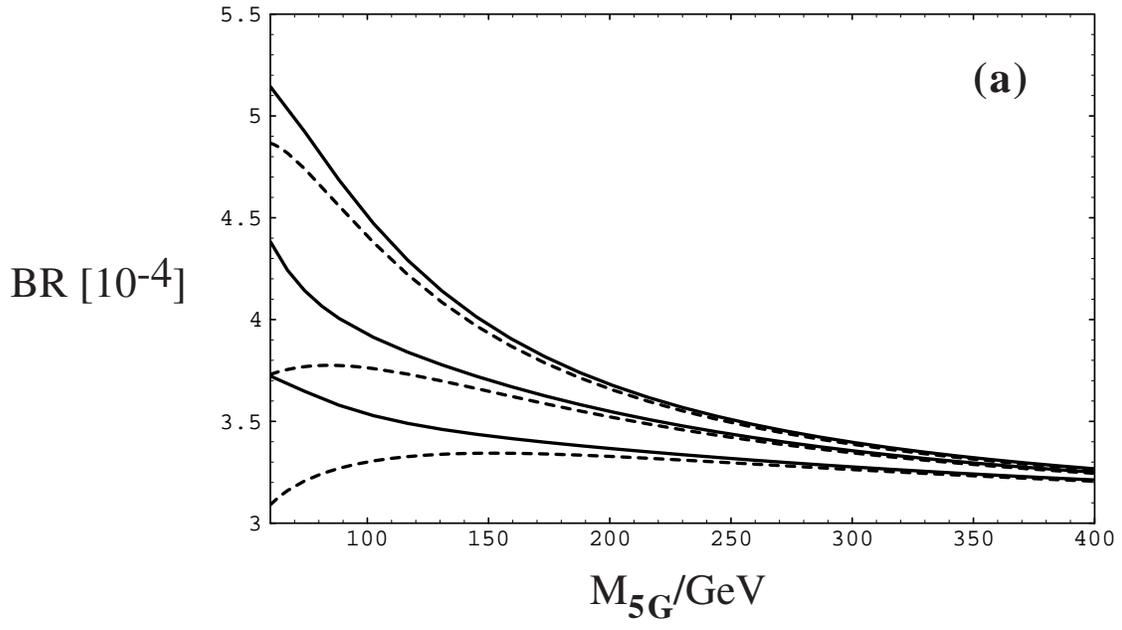
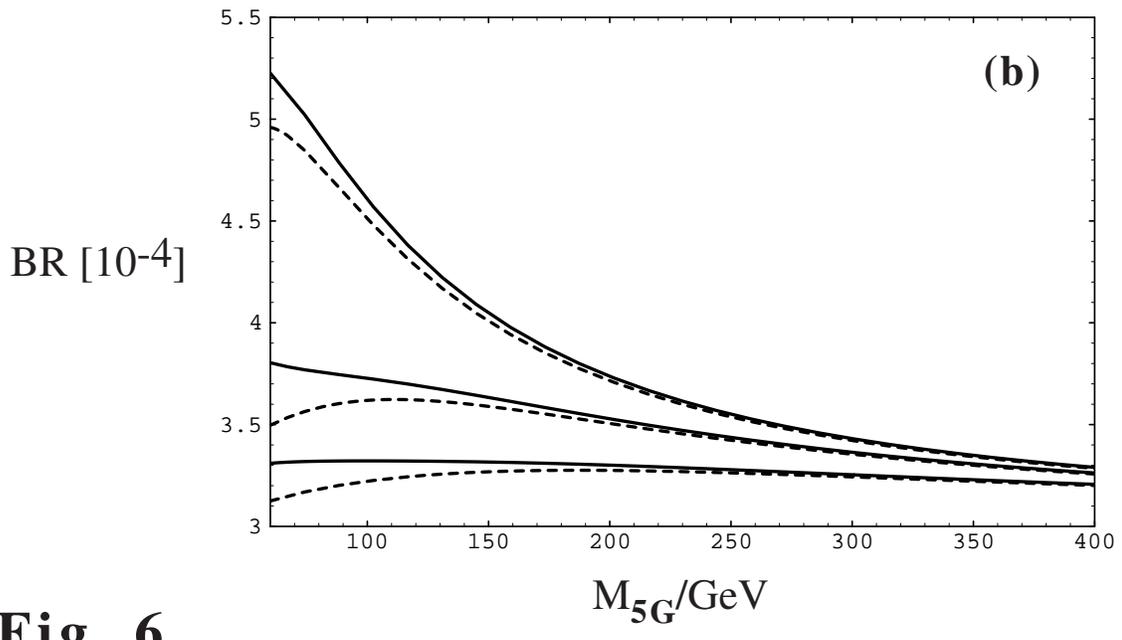
\end{document}